\documentclass{ws-procs975x65}

\begin{document}

\title{GROUND STATE PROPERTIES OF MULTI-POLARON SYSTEMS\footnote{\copyright\, 2012 by the authors. This paper may be reproduced, in its entirety, for non-commercial purposes.}}

\author{Rupert L. Frank}
\address{Department of Mathematics, Princeton University,\\
Princeton, NJ 08544, USA\\
E-mail: rlfrank@math.princeton.edu}

\author{Elliott H. Lieb}
\address{Departments of Mathematics and Physics, Princeton University,\\
Princeton, NJ 08544, USA\\
E-mail: lieb@princeton.edu}

\author{Robert Seiringer}
\address{Department of Mathematics, McGill University,\\
805 Sherbrooke Street West, Montreal, QC H3A 2K6, Canada\\
E-mail: robert.seiringer@mcgill.ca}

\author{Lawrence E. Thomas}
\address{Department of Mathematics, University of Virginia,\\
Charlottesville, VA 22904, USA\\
E-mail: let@virginia.edu}

\begin{abstract}
We summarize our recent results on the ground state energy of multi-polaron systems. In particular, we discuss stability and existence of the thermodynamic limit, and we discuss the absence of binding in the case of large Coulomb repulsion and the corresponding binding--unbinding transition. We also consider the Pekar-Tomasevich approximation to the ground state energy and we study radial symmetry of the ground state density.
\end{abstract}


\bodymatter

\section{The Fr\"ohlich Hamiltonian for a single polaron}

The large polaron was first considered by H. Fr\"ohlich in 1937 as a model of an electron interacting with the quantized optical modes of a polar crystal \cite{Fr}. In suitable units, its Hamiltonian is
\begin{eqnarray}\label{eq:h1}
 H^{(1)}(\alpha) =\mathbf{p}^2 -\sqrt\alpha\, \varphi(\mathbf{x}) + \int_{\mathbb{R}^3} a^\dagger(\mathbf{k})a(\mathbf{k})\,d\mathbf{k}
\end{eqnarray}
with
$$
\varphi(\mathbf{x}) = \frac{1}{\sqrt 2\, \pi} \int_{\mathbb{R}^3} \frac{1}{|\mathbf{k}|}[a(\mathbf{k})\exp(i\mathbf{k}\cdot\mathbf{x}) +h.c.] \,d\mathbf{k} \,.
$$ 
This Hamiltonian can be defined as a self-adjoint and lower semi-bounded operator in the Hilbert space $L^2(\mathbb{R}^3)\otimes \mathcal F$, where $\mathcal F$ is the bosonic Fock space over $L^2(\mathbb{R}^3)$ for the longitudinal optical modes of the crystal, with scalar creation and annihilation operators $a^\dagger(\mathbf{k})$ and $a(\mathbf{k})$ satisfying $[a(\mathbf{k}),a^\dagger(\mathbf{k}')]=\delta(\mathbf{k}-\mathbf{k}')$.
The momentum of an electron is $\mathbf{p}=-i\nabla$, and the coupling constant is $\alpha>0$. The ground state energy of $H^{(1)}(\alpha)$,
$$
E^{(1)}(\alpha) = \mathrm{inf}\,\mathrm{spec}\, H^{(1)}(\alpha) \,,
$$
has been studied in detail and we summarize the following properties.
\begin{enumerate}
 \item[(i)] For all $\alpha$ one has the upper \cite{Gu,LePi,LeLoPi} and lower \cite{LiYa} bounds
$$
-\alpha-\alpha^2/3\leq E^{(1)}(\alpha)\leq -\alpha \,.
$$
As a consequence, $E^{(1)}(\alpha)\sim -\alpha$ for $\alpha$ small.
\item[(ii)] Using a product function Pekar \cite{Pe} showed that 
$$
E^{(1)}(\alpha)\leq - C_P \alpha^2
$$
for all $\alpha$. Donsker and Varadhan \cite{DoVa} showed that this bound is asymptotically correct and Lieb and Thomas \cite{LiTh} obtained the error estimate
$$
E^{(1)}(\alpha)\geq - C_P \alpha^2 - \mathrm{const}\, \alpha^{9/5}
$$
for large $\alpha$. Here, $C_P=0.109$ is the number determined by Pekar's variational problem for the electron density\cite{Mi},
\begin{eqnarray}\label{eq:pekar}
C_P = \inf\left\{ \int_{\mathbb{R}^{3}} |\nabla \psi|^2 \,d\mathbf{x} - \iint_{\mathbb{R}^3\times\mathbb{R}^3} \frac{|\psi(\mathbf{x})|^2\, |\psi(\mathbf{y})|^2}{|\mathbf{x}-\mathbf{y}|} \,d\mathbf{x}\,d\mathbf{y} : \|\psi\|_2 =1  \right\} \,.
\end{eqnarray}
The minimizing $\psi$ is unique up to translations and a trivial phase \cite{Li}.
\item[(iii)] There is a representation for $E^{(1)}(\alpha)$ in
terms of path integrals \cite{Fe}.  In terms of the partition function
$Z_T^{(1)}(\alpha)=\mathrm{Tr}\, \exp\big(-T H^{(1)}(\alpha)\big)$, one has $E^{(1)}(\alpha) =
-\lim_{T\to\infty}T^{-1}\log Z_T^{(1)}(\alpha)$.  (Strictly speaking,
$Z_T^{(1)}(\alpha)$ does not exist because of the translation invariance of
$H^{(1)}(\alpha)$ and the infinite number of phonon modes. These
technicalities can be handled by inserting appropriate cutoffs, to be
removed at the end of the calculation \cite{Ro,Sp}.) After one integrates out the phonon variables, $Z_T^{(1)}(\alpha)$ has the functional integral representation
\begin{equation}
 \label{eq:trace}
Z_T^{(1)}(\alpha) = \int d\mu \exp\left[ \frac\alpha2 \int_0^T\!\! \int_0^T \!\frac{e^{-|t-s|}\,dt\,ds}{|\mathbf{x}(t)-\mathbf{x}(s)|} \right] \,,
\end{equation}
where $d\mu$ is Wiener measure on all $T$-periodic paths $\mathbf{x}(t)$. (In physics notation $d\mu=\exp(-\int_0^T \dot{\mathbf{x}}(t)^2 \,dt)\, d\,\mathrm{path}$. Strictly speaking, $t-s$ has to be understood modulo $T$, but this is irrelevant as $T\to \infty$.)
\end{enumerate}


\section{Multi-polaron systems and their thermodynamic stability}

The Hamiltonian for $N$ polarons is
\begin{eqnarray}\label{eq:ham}
 H^{(N)}_U(\alpha) = & \sum_{j=1}^N \left( \mathbf{p}_{j}^2 -\sqrt\alpha\, \varphi(\mathbf{x}_j) \right) + \int_{\mathbb{R}^3} a^\dagger(\mathbf{k})a(\mathbf{k})\,d\mathbf{k} +U \sum_{1\leq i<j\leq N} \frac1{|\mathbf{x}_i -\mathbf{x}_j|}
\end{eqnarray}
and we denote its ground state energy by
$$
E^{(N)}_U(\alpha) = \mathrm{inf}\,\mathrm{spec}\, H^{(N)}_U(\alpha) \,.
$$
This operator acts in the Hilbert space $L^2(\mathbb{R}^{3N})\otimes\mathcal F$. We ignore Fermi statistics for the electrons, because its imposition changes things only quantitatively, not qualitatively. The Coulomb repulsion parameter $U\geq 0$ is equal to $e^2$. Fr\"ohlich's derivation \cite{Fr} of $H^{(N)}_U(\alpha)$ implies that $U>2\alpha$,  and this is crucial for thermodynamic stability, as we shall see.

We consider the question of the existence of the thermodynamic
limit for a multi-polaron system in the ground state. For large $N$, physical intuition suggests that $E^{(N)}_U(\alpha)\sim
-\mathrm{const}\, N$. This supposition is known to be \emph{false} if $U<2\alpha$.
Indeed \cite{GrMo}, even with the Pauli
principle, $E^{(N)}_U(\alpha)\sim -\mathrm{const}\, N^{7/3}$ when $U<2\alpha$. Absent
the Pauli principle, $E^{(N)}_U(\alpha)$ would behave even worse, as $-\mathrm{const}\,
N^3$. It is also known \cite{GrMo} that $E^{(N)}_U(\alpha)\geq -\mathrm{const}\, N^{2}$
if $U>2\alpha$.  The latter bound ought to be $-\mathrm{const}\, N$ instead, and this is indeed the statement of the following theorem \cite{FrLiSeTh0,FrLiSeTh}.

\begin{theorem}[\textbf{Thermodynamic stability for {\mathversion{bold} $U>2\alpha$}}] \label{thm:stabalpha}
 For any $U>2\alpha>0$, there is a constant $C(U,\alpha)$ such that for all $N\geq 2$,
$$
E_U^{(N)}(\alpha) \geq - C(U,\alpha)\ N \,.
$$
\end{theorem}

Our lower bound on $N^{-1} E^{(N)}_U(\alpha)$ goes to $-\infty$ as
$U\searrow2\alpha$, but we are not claiming that this reflects the
true state of affairs. Whether $\lim_{N\to\infty} N^{-1}
E^{(N)}_{2\alpha}(\alpha)$ is finite or not remains an \textit{open
  problem}. There are results on this question in the Pekar-Tomasevich approximation \cite{GrMo,BeFrLi}.

The linear lower bound from Theorem \ref{thm:stabalpha}, together with the sub-additivity of the energy \cite{GrMo,LiSe}, i.e., 
\begin{equation}\label{eq:subadd}
E^{(N+M)}_U(\alpha) \leq E^{(N)}_U(\alpha)+ E^{(M)}_U(\alpha)\,,
\end{equation}
implies:

\begin{corollary}[\textbf{Thermodynamic limit for {\mathversion{bold} $U>2\alpha$}}] \label{thm:tlalpha}
 For any $U>2\alpha>0$, $\lim_{N\to\infty} N^{-1} E^{(N)}_U(\alpha)$ exists.
\end{corollary}


\section{Binding and non-binding of multi-polaron systems}

The binding of polarons, or its absence, is an old and subtle topic. For some time the bipolaron binding energy $\Delta E_U(\alpha)= 2 E^{(1)}(\alpha)-E^{(2)}_U(\alpha)$ was thought to be zero for all $U\geq 2\alpha$, on the basis of an 
inadequate variational calculation, but it is now known \cite{DePeVe}
to be positive for some $U>2\alpha$. The first question we address is
whether $\Delta E_U(\alpha)=0$ for $U$ sufficiently large.
It is understood that the effective interaction induced by the phonon
field for two polarons at large distances $d$ is approximately
Coulomb-like $-2\alpha/d$, but this alone does not preclude binding. (The reason for $2\alpha\cdot \mathrm{distance}^{-1}$ can be seen from the $N$-polaron analogue of \eqref{eq:trace},
\begin{equation*}
 Z^{(N)}_{T,U}(\alpha) \!= \!\!\int \!d\mu^{(N)} \exp\!\Big( \frac\alpha2 \sum_{i,j} \int_0^T\!\!\! \int_0^T \!\!\frac{e^{-|t-s|}\,dt\,ds}{|\mathbf{x}_i(t)-\mathbf{x}_j(s)|} 
- U \sum_{i<j} \int_0^T \!\!\!\frac{dt}{|\mathbf{x}_i(t) -\mathbf{x}_j(t)|} \Big),
\end{equation*}
where $d\mu^{(N)}$ is Wiener measure on all $T$-periodic paths $(\mathbf{x}_1(t),\ldots,\mathbf{x}_N(t))$. There is a factor $\alpha/2$, but the pair $(i,j)$ appears twice, and the integral $\int_\mathbb{R} e^{-|t-s|}\,ds = 2$.)
The known existence of bipolarons for some $U>2\alpha$ is an effect of
correlations. It is \emph{a priori} conceivable that correlations lead to an
effective attraction that is stronger than Coulomb at large distances.
If it were, for example, equal to $(2\alpha/d)\log(\log(\log(d)))$,
then this minuscule perturbation of Coulomb's law, which would be
virtually undetectable by a variational calculation, would result in
binding for \emph{all} $U$. The absence of binding is a problem that
has resisted a definitive resolution for many years. We proved \cite{FrLiSeTh0,FrLiSeTh}:

\begin{theorem}[\textbf{Absence of binding for {\mathversion{bold}$N$} polarons}]\label{thm:nobindingN}
 For any $\alpha>0$ there is a $U_c(\alpha)<\infty$ such that for all $U\geq U_c(\alpha)$ and all $N\geq 2$ one has
\begin{equation}
 \label{eq:bindabs}
E^{(N)}_U(\alpha) = N E^{(1)}(\alpha) \,.
\end{equation}
\end{theorem}

In particular, for $N=2$ we show that $E^{(2)}_U(\alpha) = 2 E^{(1)}(\alpha)$ provided $U\geq 2C\alpha$ with $C=26.6$. The constant $26.6$ vastly exceeds the current, non-rigorous estimates of about $1.15$ \cite{VSPD,fomin}, so it is an \emph{open problem} to find a more accurate rigorous bound. Somewhat better bounds are known in the Pekar--Tomasevich approximation \cite{FrLiSeTh,BeBl}.

While our bound for $U_c(\alpha)$ is linear in $\alpha$ for large $\alpha$, we have not achieved this linear bound for small $\alpha$ and this remains an \textit{open problem}.

Theorem \ref{thm:nobindingN} says that
$$
U_c^{(N)}(\alpha)=\inf\left\{ U \geq 0:\ E^{(N)}_{U'}(\alpha) = N E^{(1)}(\alpha) \ \text{for all}\ U'\geq U \right\}
$$
is finite and bounded uniformly in $N$. For any $U> U_c^{(N)}(\alpha)$ and any state $\Psi$
\begin{equation}
 \label{eq:energybound}
\left\langle \Psi \left| H_U^{(N)}(\alpha) \right|\Psi\right\rangle \geq N E^{(1)}(\alpha) \|\Psi\|^2 + \left(U-U_c^{(N)}(\alpha)\right) \ \Big\langle \Psi \Big|  \sum_{i<j} |\mathbf{x}_i-\mathbf{x}_j|^{-1} \Big|\Psi \Big\rangle \,.
\end{equation}
This is a quantitative estimate of the energy penalty needed to bring two or more particles within a finite distance of each other. In particular, it implies that for $U>U_c^{(N)}(\alpha)$ there cannot be a normalizable ground state, even in a fixed momentum sector. Inequality \eqref{eq:energybound} is not only true for our bound on $U_c(\alpha)$, but also for the (unknown) exact value of $U_c^{(N)}(\alpha)$.

For $U$ in the range $2\alpha<U<U_c(\alpha)$, there are bound states
of an undetermined nature. Does the system become a gas of bipolarons, or does it coalesce into a true $N$-particle bound state? If the latter, does this state exhibit a periodic structure, thereby forming a super-crystal on top of the underlying lattice of atoms? This is perhaps the physically most interesting \textit{open problem}. While particle statistics does not play any role for our main results, the answer to this question will crucially depend on particle statistics (Bose or Fermi) \cite{SVPD1,SVPD2}.


\section{Binding--unbinding transition}

We now discuss the behavior of the $N$-polaron radius as the repulsion parameter $U$ approaches from within the binding regime a critical value where $N$-polarons cease to be bound. Does the $N$-polaron radius in this limit increase towards infinity or does it remain finite? Verbist, Peeters and Devreese \cite{VPD} proposed a `first-order' transition, that is, the Coulomb repulsion jumps discontinuously from a positive value to zero and the radius, too, jumps discontinuously. We prove this rigorously \cite{FrLiSe1} under the assumption that the critical value is strictly bigger than $2\alpha$. This is known to be satisfied for large $\alpha$ \cite{Lw}.

Besides the ground state energy $E^{(N)}_U(\alpha)$ of the Hamiltonian $H^{(N)}_U(\alpha)$ we also need the \emph{minimum break-up energy}
\begin{equation}
 \label{eq:mbe}
\widetilde E^{(N)}_U(\alpha) = \min_{1\leq n\leq N-1}
\left( E^{(n)}_U(\alpha) + E^{(N-n)}_U(\alpha) \right)\,.
\end{equation}
Note that it is always the case that $E^{(N)}_U(\alpha)\leq \widetilde E^{(N)}_U(\alpha)$.

\begin{theorem}[\textbf{Upper bound on the $N$-polaron
radius}]\label{thm:bindingN}
For any $N\geq 2$ and  $\epsilon>0$ there is a constant $C_\epsilon(N)>0$  such that for all
$0< 2\alpha(1+\epsilon) < U$ with $ E^{(N)}_U(\alpha) < \widetilde E_U^{(N)}(\alpha)$
and all states $\Psi$
\begin{equation} \label{eq:bindabs2} \left\langle \Psi \left| \frac
      1{\max_{i\neq j}|\mathbf{x}_i-\mathbf{x}_j|} \right| \Psi \right \rangle \geq
  \frac { U - 2 \alpha(1+\epsilon)}{C_\epsilon(N) ( 1 + U /\alpha)}
  \frac{ \left\langle \Psi \left| \widetilde E^{(N)}_U(\alpha) -
        H^{(N)}_U(\alpha) \right| \Psi \right \rangle}{\widetilde
    E^{(N)}_U(\alpha) - E_U^{(N)}(\alpha)} \,.
\end{equation}
\end{theorem}

Since (\ref{eq:bindabs2}) holds for all $\Psi$, it can be
reformulated as an operator inequality. The bound is
non-trivial only for states
$\Psi$ with $\langle\Psi|H^{(N)}_U(\alpha)|\Psi\rangle < \widetilde
E^{(N)}_U(\alpha) \|\Psi\|^2$, however, which exist since $ E^{(N)}_U(\alpha) < \widetilde
E_U^{(N)}(\alpha)$ by assumption. For approximate ground states, that is, states satisfying
$$
\langle\Psi|H^{(N)}_U(\alpha)|\Psi\rangle \leq (1-\theta) \widetilde
E^{(N)}_U(\alpha) \|\Psi\|^2+ \theta E^{(N)}_U(\alpha)\|\Psi\|^2
$$
for some $\theta >0$, (\ref{eq:bindabs2}) gives the uniform lower bound
$$
\left\langle \Psi \left| \frac
      1{\max_{i\neq j}|\mathbf{x}_i-\mathbf{x}_j|} \right| \Psi \right \rangle \geq
  \frac { U - 2 \alpha(1+\epsilon)}{C_\epsilon(N) ( 1 + U /\alpha)}
  \ \theta \|\Psi\|^2 \,,
$$
which means a uniform upper bound on the radius of the multipolaron system. This bound depends only on the value
of $\theta$ and does not explode as $U$ approaches the critical unbinding value.

A similar phenomenon was shown by T. and M. Hoffmann-Ostenhof and B.~Simon \cite{HoHoSi} for a two-electron atom and the proof strategy of Theorem \ref{thm:bindingN} applies to that problem as well \cite{FrLiSe1}. The overall lesson is that this kind of discontinuous binding will occur whenever the net repulsion at large distances falls of slower than $r^{-2}$.


\section{The Pekar--Tomasevich approximation}

The Pekar--Tomasevich approximation to the ground state energy $E_U^{(N)}(\alpha)$ consists in minimizing $\langle\Psi |H^{(N)}_U(\alpha) |\Psi\rangle$ only over $\Psi$'s of the form $\psi\otimes \Phi$, where $\psi\in L^2(\mathbb{R}^{3N})$, $\Phi\in\mathcal F$, and both $\psi$ and $\Phi$ are normalized. For $N=1$ we obtain functional \eqref{eq:pekar}. In the $N$-polaron case this approximation leads to the minimization of the following Pekar--Tomasevich functional for normalized functions $\psi$ on $\mathbb{R}^{3N}$,
\begin{eqnarray*}\nonumber
\sum_{i=1}^N \int_{\mathbb{R}^{3N}} |\nabla_i \psi|^2 \,dX + U \sum_{i<j} \int_{\mathbb{R}^{3N}} \frac{|\psi(X)|^2}{|\mathbf{x}_i-\mathbf{x}_j|} \,dX 
- \alpha \iint_{\mathbb{R}^3\times\mathbb{R}^3} \frac{\rho_\psi(\mathbf{x})\, \rho_\psi(\mathbf{y})}{|\mathbf{x}-\mathbf{y}|} \,d\mathbf{x}\,d\mathbf{y} \,,
\end{eqnarray*}
where $dX=\prod_{k=1}^N d\mathbf{x}_k$. The density $\rho_\psi$ of $\psi$ is defined as usual by
\begin{equation*}
\rho_\psi(\mathbf{x}) = \sum_{i=1}^N \int_{\mathbb{R}^{3(N-1)}} |\psi(\mathbf{x}_1,\ldots,\mathbf{x},\ldots,\mathbf{x}_N)|^2 \,d\mathbf{x}_1\cdots \widehat{d\mathbf{x}_i} \cdots d\mathbf{x}_N
\end{equation*}
with $\mathbf{x}$ at the $i$-th position, and $\widehat{d\mathbf{x}_i}$ meaning that $d\mathbf{x}_i$ has to be omitted in the product $\prod_{k=1}^N d\mathbf{x}_k$. Since the Pekar--Tomasevich functional is the result of a variational calculation, its energy gives an upper bound to the ground state energy $E^{(N)}_U(\alpha)$.

The Pekar--Tomasevich minimization problem has been studied in great detail \cite{Li,MiSp,GrMo,FrLiSeTh,Lw,BeBl,FrLiSe1,FrLiSe2,BeFrLi}. In particular, the analogous statements of our Theorems \ref{thm:stabalpha}, \ref{thm:nobindingN} and \ref{thm:bindingN} remain valid for this model.

Here, for the sake of simplicity, we restrict ourselves to the case of a bipolaron, so that the energy functional becomes
\begin{eqnarray*}
\mathcal E_U[\psi] = & \iint_{\mathbb{R}^3\times\mathbb{R}^3} \left( |\nabla_\mathbf{x}\psi|^2 +|\nabla_\mathbf{y}\psi|^2 + \frac{U}{|\mathbf{x}-\mathbf{y}|} |\psi|^2 \right) \,d\mathbf{x}\,d\mathbf{y} \\
& - \alpha \iint_{\mathbb{R}^3\times\mathbb{R}^3} \frac{\rho_\psi(\mathbf{x})\, \rho_\psi(\mathbf{y})}{|\mathbf{x}-\mathbf{y}|} \,d\mathbf{x}\,d\mathbf{y} \,.
\end{eqnarray*}
The electron coordinates are $\mathbf{x}$ and $\mathbf{y}$ and the electron spin does not appear explicitly, except that $\psi$ is symmetric for the ground state, which is a singlet state.

There is a considerable literature on the subject of rotation invariance of the bipolaron energy minimizer, usually formulated in the language of `one-center bipolaron versus two-center bipolaron' \cite{SaMi,KaLaSy,KaLaSy2,KaLa}. The analyses are all based on variational calculations. While there seems to be general agreement that the one-center bipolaron has the lower energy, it is not completely clear that a more sophisticated variational treatment will preserve rotational symmetry, especially near the value of $U$ where the bipolaron ceases to be bound.

The value of $U$ determined by physical electrostatic considerations is always $U>2\alpha$. Nevertheless, one can consider the mathematical question for small, but positive $U$ and ask whether there is a possible lack of rotational invariance in that case. After all, a rotating object like the earth becomes oblate even for the smallest amount of rotation. 

Our theorem \cite{FrLiSe2} says that for the bipolaron rotational symmetry is not broken for small $U$:

\begin{theorem}\label{thm:symm}
 There is a $\nu_s>0$ such that for all $U < \nu_s \alpha$ the minimizer of $\mathcal E_U$ is unique up to translations and multiplication by a constant phase. In particular, after a translation it is rotation invariant, that is, $\psi(\mathcal R \mathbf{x},\mathcal R \mathbf{y})=\psi(\mathbf{x},\mathbf{y})$ for any $\mathbf{x},\mathbf{y}\in\mathbb{R}^3$ and any $\mathcal R\in O(3)$.
\end{theorem}

Our proof of Theorem \ref{thm:symm} is perturbative in nature. It uses crucially a non-degeneracy statement about the single polaron functional \eqref{eq:pekar} due to Lenzmann \cite{Le}.

We also show that for certain values of $U/\alpha$ the bipolaron equation has a positive solution which is \emph{not} a minimizer.

It remains an \emph{open problem} to decide whether the ground state ceases to be rotation invariant for $U$ close to the critical value where bipolarons cease to be bound in the Pekar--Tomasevich approximation.


\section*{Acknowledgments}
 Partial financial support from the U.S.~National Science Foundation through grants PHY-1068285 (R.F.), PHY-0965859 (E.L.), the Simons Foundation (\# 230207, E.L.) and the NSERC (R.S.) is acknowledged. L.T. would like to thank the PIMS Institute, University of British Columbia, for their hospitality and support. We are grateful to Herbert Spohn for stimulating our interest.


\begin{thebibliography}{35}
\bibitem{BeBl} R. D. Benguria, G. A. Bley, \textit{Exact asymptotic behavior of the Pekar--Tomasevich functional}. J. Math. Phys. \textbf{52} (2011), no. 5, 052110.
\bibitem{BeFrLi} R. D. Benguria, R. L. Frank, E. H. Lieb, in preparation.
\bibitem{DePeVe} J. T. Devreese, F. M. Peeters, G. Verbist, \textit{Large bipolarons in two and three dimensions}. Phys. Rev. {\bf B 43} (1991), 2712--2720.
\bibitem{DoVa} M. Donsker, S. R. S. Varadhan, \textit{Asymptotics for the polaron}. Comm. Pure Appl. Math. \textbf{36} (1983), 505--528.
\bibitem{Fe} R. P. Feynman, \textit{Slow electrons in a polar crystal}. Phys. Rev. \textbf{97} (1955), 660--665.
\bibitem{FrLiSeTh0} R. L. Frank, E. H. Lieb, R. Seiringer, L. E. Thomas, \textit{Bi-polaron and N-polaron binding energies}. Phys. Rev. Lett. \textbf{104} (2010), 210402.
\bibitem{FrLiSeTh} R. L. Frank, E. H. Lieb, R. Seiringer, L. Thomas, \textit{Stability and absence of binding for multi-polaron systems}. Publ. Math. IHES \textbf{113} (2011), no. 1, 39--67.
\bibitem{FrLiSe1} R. L. Frank, E. H. Lieb, R. Seiringer, \textit{Binding of polarons and atoms at threshold}. Comm. Math. Phys. \textbf{313} (2012), no. 2, 405--424.
\bibitem{FrLiSe2} R. L. Frank, E. H. Lieb, R. Seiringer, \textit{Symmetry of bipolaron bound states for small Coulomb repulsion}. Comm. Math. Phys., to appear. Preprint (2012): arXiv:1201.3954.
\bibitem{Fr} H. Fr\"ohlich, \textit{Theory of electrical breakdown in ionic crystals}. Proc. R. Soc. Lond. A \textbf{160} (1937), 230--241.
\bibitem{GrMo} M. Griesemer, J. Schach M\o ller, \textit{Bounds on the minimal energy of translation invariant $N$-polaron systems}.  Comm. Math. Phys.  \textbf{297}  (2010),  no. 1, 283--297.
\bibitem{Gu} M. Gurari, \textit{Self-energy of slow electrons in polar materials}. Phil. Mag. Ser. 7 \textbf{44}:350 (1953), 329--336.
\bibitem{HoHoSi} M. Hoffmann-Ostenhof, T. Hoffmann-Ostenhof, B. Simon, {\it A
multiparticle Coulomb system with bound state at threshold}, J. Phys. A {\bf
16}, 1125--1131 (1983).
\bibitem{KaLaSy} N. I. Kashirina, V. D. Lakhno, V. V. Sychev, \textit{Electron correlations and instability of a two-center bipolaron}. Phys. Solid State \textbf{45} (2003), no. 1, 171--175.
\bibitem{KaLaSy2} N. I. Kashirina, V. D. Lakhno, V. V. Sychyov, \textit{Polaron effects and electron correlations in two-electron systems: Arbitrary value of electron-phonon interaction}. Phys. Rev. B \textbf{71} (2005), 134301.
\bibitem{KaLa} N. I. Kashirina, V. D. Lakhno, \textit{Large-radius bipolaron and the polaron-polaron interaction}. Phys. Usp. \textbf{53} (2010), no. 5, 431--453.
\bibitem{LePi} T-D. Lee, D. Pines, \textit{The motion of slow electrons in polar crystals}. Phys. Rev. \textbf{88} (1952), 960--961.
\bibitem{LeLoPi} T. D. Lee, F. Low, D. Pines, \textit{The motion of slow electrons in a polar crystal}. Phys. Rev. \textbf{90} (1953), 297--302.
\bibitem{Le} E. Lenzmann, \textit{Uniqueness of ground states for pseudorelativistic Hartree equations}. Anal. PDE \textbf{2} (2009), no. 1, 1--27. 
\bibitem{Lw} M. Lewin, {\it Geometric methods for nonlinear many-body quantum systems}, J. Funct. Anal. {\bf 260} (2011), 3535--3595.
\bibitem{Li} E. H. Lieb, \textit{Existence and uniqueness of the minimizing solution of Choquard's nonlinear equation}.
Studies in Appl. Math. \textbf{57} (1976/77), no. 2, 93--105.
\bibitem{LiSe} E. H. Lieb, R. Seiringer, \textit{The stability of matter in quantum mechanics}, Cambridge (2010).
\bibitem{LiTh} E. H. Lieb, L. E. Thomas, \textit{Exact ground state energy of the strong-coupling polaron}. Comm. Math. Phys. \textbf{183} (1997), no. 3, 511--519. Erratum: \textit{ibid.} \textbf{188} (1997),  no. 2, 499--500.
\bibitem{LiYa} E. H. Lieb, K. Yamazaki, \textit{Ground-state energy and effective mass of the polaron}. Phys. Rev. \textbf{111} (1958), 728--722.
\bibitem{MiSp} T. Miyao, H. Spohn, {\it The bipolaron in the strong coupling limit}. Ann. Henri Poincar{\'e} {\bf 8} (2007), 1333--1370.
\bibitem{Mi} S. J. Miyake, \textit{Strong coupling limit of the polaron ground state}. J. Phys. Soc. Jpn. \textbf{38} (1975), 181--182.
\bibitem{Pe} S. I. Pekar, \textit{Untersuchung \"uber die Elektronentheorie der Kristalle}, Berlin, Akad. Verlag (1954).
\bibitem{Ro} G. Roepstorff, \textit{Path integral approach to quantum phy\-sics}. Berlin-Heidelberg-New York, Springer, 1994.
\bibitem{SaMi} S. Sahoo, T. K. Mitra, \textit{Molecular-orbital approach to the Fr\"ohlich bipolaron}, Phys. Rev. B \textbf{48} (1993), no. 9, 6019--6023.
\bibitem{fomin} M. A. Smondyrev, V.M. Fomin, {\it Pekar-Fr\"ohlich bipolarons}. In: {\it Polarons and applications}, Proceedings in Nonlinear Science, V.D. Lakhno, ed., Wiley (1994).
\bibitem{SVPD2} M. A. Smondyrev, A. A. Shanenko,  J. T. Devreese, \textit{Stability criterion for large bipolarons in a polaron-gas background}, Phys. Rev. B {\bf 63} (2000), 024302.
\bibitem{SVPD1} M. A. Smondyrev, G. Verbist, F. M. Peeters, J. T. Devreese, \textit{Stability of multipolaron matter}, Phys. Rev. B  {\bf 47} (1993),  2596--2601.
\bibitem{Sp} H. Spohn, \textit{The polaron functional integral}. In: Stochastic processes and their applications, Dordrecht-Boston-London, Kluwer, 1990.
\bibitem{VPD} G. Verbist, F. M. Peeters, J. T. Devreese, \textit{Large bipolarons in two and three dimensions}. Phys. Rev. B {\bf 43} (1991), 2712--2720.
\bibitem{VSPD} G. Verbist,  M. A. Smondyrev, F. M. Peeters, J. T. Devreese,
\textit{Strong-coupling analysis of large bipolarons in two  and three  dimensions}, Phys. Rev. B {\bf 45} (1992), 5262--5269.

\end{thebibliography}

\end{document}